\documentclass[prx,twocolumn,showpacs,superscriptaddress]{revtex4-1}

\usepackage{multirow,eurosym,amssymb,amsfonts,amsmath,setspace,graphicx,color,bm,float,verbatim}

\newcommand{\bra}[1]{\langle #1|}
\newcommand{\ket}[1]{|#1\rangle}

\def\ket#1{| #1 \rangle}
\def\bra#1{\langle #1 |}

\def\be{\begin{equation}}
\def\ee{\end{equation}}

\def\bsplit{\begin{split}}
\def\nsplit{\end{split}}

\begin{document}
\title{Hybrid teleportation via entangled coherent states in circuit quantum electrodynamics}

\date{\today}

\author{Jaewoo Joo}
\affiliation{Advanced Technology Institute and Department of Physics, University of Surrey, Guildford, GU2 7XH, United Kingdom}

\author{Eran Ginossar}
\affiliation{Advanced Technology Institute and Department of Physics, University of Surrey, Guildford, GU2 7XH, United Kingdom}

\begin{abstract}
We propose a deterministic scheme for teleporting an unknown qubit through continuous-variable entangled states in superconducting circuits. The qubit is a superconducting two-level system and the bipartite quantum channel is a photonic entangled coherent state between two cavities. A Bell-type measurement performed on the hybrid state of solid and photonic states brings a discrete-variable unknown electronic state to a continuous-variable photonic cat state in a cavity mode. This scheme further enables applications for quantum information processing in the same architecture of circuit-QED such as verification and error-detection schemes for entangled coherent states. Finally, a dynamical method of a self-Kerr tunability in a cavity state has been investigated for minimizing self-Kerr distortion and all essential ingredients are shown to be experimentally feasible with the state of the art superconducting circuits. 
\end{abstract}


\maketitle
\section{Introduction}
The scheme of quantum teleportation \cite{original} is of the essence for technological applications of quantum information processing such as quantum cryptography in multipartite quantum networks \cite{Quan_Crypto} and for measurement based quantum computation \cite{MBQC}. In the original teleportation scheme, an unknown qubit from a sender (Alice) can be deterministically teleported to a receiver (Bob) by performing Bell-state measurement (BSM) through a bipartite entangled state (called a channel) in discrete variables (DVs). After the feed-forward of classical information, one can recover the original qubit state at the other location of the channel. Since the scheme of postselected DV teleportation has been firstly demonstrated in quantum optics \cite{Bouwmeister, Photonic_tele}, teleportation schemes have also been demonstrated in other physical systems, particularly for deterministic methods in ion traps \cite{Atomic_tele}, atomic ensembles \cite{Polzik} and superconducting circuits \cite{qubit-teleport13}. 

An alternative representation, called continuous-variable (CV) quantum teleportation \cite{Vaidman}, has been in parallel investigated because a CV channel is indeed a natural resource for entanglement (e.g., a position-momentum entangled state in the Einstein-Podolsky-Rosen's paper \cite{EPR}). For example, the first demonstration of unconditional teleportation has been successfully performed in nonclassical CV states (e.g., using two-mode squeezed states) \cite{Sam_Braunstein}. CV teleportation is essential to the schemes of CV quantum information processing which have shown advantages as compared with DV-qubit information processing  \cite{CV_RMP, compare_CV_tele} (e.g., time-frequency encoding \cite{Time_Freq}, fault-tolerant CV quantum computing \cite{CV-MBQC,cat_QC_threshold}). 

One of the CV-qubit representations is based on Schr\"odinger cat states (SCS) \cite{Schro01} given by the superposition of two phase-opposite coherent states \cite{Peter05}. The SCSs have been created in various methods (e.g., photon adding and subtracting schemes in quantum optics \cite{Grangier} and ion- or Rydberg atom-cavity systems \cite{Haroche,Wineland}) and relatively larger SCSs have been very recently achieved in circuit-QED \cite{Yale_big_cat}. A CV qubit can in principle encode information beyond DV qubits because it is described in infinite dimension \cite{Sam_Braunstein, Jeong02, Sanders92, CVentanglement}. For example, a generalized SCS with many different phases can be used to realise a qu$d$it which will be of use for hardware-efficient quantum memory \cite{Cat_QEC}. Thus, DV-CV hybrid teleportation is not only an alternative for DV-qubit teleportation but also advantageous for practical quantum information processing \cite{Jeong15}. For instance, these recent developments will lead innovative tools for measurement-based quantum computing using hybrid single- and two-qubit gates \cite{Andersen+Akira15}.

Here we develop a DV-CV hybrid teleportation scheme specifically designed to be implemented on a superconducting circuit. It is a key building block required for measurement-based quantum computation\cite{MBQC,cat_QC_threshold,Ralph11} because a series of teleportations can mimick one- and two-qubit gates. The scheme is physically hybrid in the sense that it teleports quantum information from a solid-state qubit to a microwave photon. An unknown qubit is prepared in a two-level superconducting qubit and an entangled coherent states (ECS), known as an excellent resource for quantum metrology and other quantum information processing \cite{JooPRL11,Sanders12,Ralph03,GrangierECS}, is created in microwave photons inside two cavities with the help of an adjacent superconducting qubit \cite{qcMAP13}. This architecture is feasible in the state-of-the-art superconducting setup used for creating a microwave SCS as shown recently in \cite{Yale_big_cat}. It consists of two cavities coupled to three superconducting qubits and additional two readout resonators. The unknown state is encoded in a superconducting DV qubit and teleported into a CV state in one of the cavities. In contrast to DV- and CV-only teleportations, we find that the teleportation fidelity depends not only on the amount of decoherence but also on the amplitude size of the ECS channel state. 

The hybrid scheme discussed here has twofold meaning: hybrid qubits imply the combination of DV and CV encoding on one hand and of hybrid quantum systems in microwave and superconducting  states on the other hand. The creation of entanglement in photonic hybrid qubits has been very recently demonstrated in quantum optics \cite{Jeong14} and optical hybrid CV teleportation has been very recently performed \cite{Furusawa13}, however, two-fold hybrid CV quantum information processing has not been well developed yet because it is difficult to create the entangled states deterministically created by non-linear optical amplifiers with a low efficiency probabilistically. Thus, the microwave ECS  with high fidelity in cavities, which can be naturally entangled with superconducting qubits, will enable to be utilised for hybrid quantum information processing practically as we investigate the verification of ECSs, single-qubit gates on ECSs, and reduction of self-Kerr effects in a cavity in Sections \ref{III} and \ref{self-Kerr}.

\section{Hybrid CV teleportation}
First, let us briefly describe quantum teleportation in DV qubits. We have an unknown qubit state in mode $A$ given by
\begin{eqnarray}
\ket{\psi}_A = a \ket{0}_A + b \ket{1}_A,
\label{Back_01} 
\end{eqnarray}
where $|a|^2 + |b|^2 = 1$. Additionally, Alice and Bob have already shared one of the Bell states in modes $B$ and $C$ such as $\ket{\Phi^{\pm}} =(\ket{00}\pm \ket{11})/ \sqrt{2}$ and $\ket{\Psi^{\pm}} = (\ket{01} \pm \ket{10})/ \sqrt{2}$ and we assume that the total state is initially prepared with $\ket{\Phi^{+}}_{BC}$ given by
\begin{eqnarray}
\ket{\Psi^{tot}}_{ABC} &=& \ket{\psi}_A \ket{\Phi^{+}}_{BC}, \\
&=& {1 \over 2} \Big( \ket{\Phi^{+}}_{AB} \otimes \openone  + \ket{\Phi^{-}}_{AB} \otimes \sigma^{z}_C \nonumber \\
&& + \ket{\Psi^{+}}_{AB} \otimes \sigma^{x}_C + \ket{\Psi^{-}}_{AB} \otimes \sigma^{x}_C \sigma^{z}_C \Big) \ket{\psi}_C. \nonumber 
\label{Back_03} 
\end{eqnarray}
This mathematical representation implies what are the teleported state dependent on the measurement outcomes of the BSM.
After Alice performs a BSM (known as a joint measurement between $A$ an $B$) in $\hat{M}_{AB} = \{ \ket{\Phi^{\pm}}_{AB} \bra{\Phi^{\pm}}, \ket{\Psi^{\pm}}_{AB} \bra{\Psi^{\pm}} \}$, she announces measurement outcomes to Bob to reconstruct the unknown state by applying one of four single-qubit operations ($\openone, \sigma^{x,y,z}$). 

\subsection{Protocol}
In the hybrid protocol, we follow the above teleportation protocol but the channel state is now made from CV states written by
\begin{eqnarray}
\ket{ECS^{\Phi+}_{\alpha}}_{BC} ={\cal N}^{+}_{\alpha} (\ket{\alpha}_{B} \ket{\alpha}_{C} + \ket{-\alpha}_{B} \ket{-\alpha}_{C}),
\label{Back_04} 
\end{eqnarray}
where ${\cal N}^{\pm}_{\alpha} = 1/\sqrt{2(1 \pm e^{-4|\alpha|^2})}$ is a normalisation and $\ket{\alpha} = \sum_{m=0}^{\infty} c_m \ket{m}$ ($c_m = e^{-|\alpha|^2/2} \alpha^m / \sqrt{m!}$). Four Bell-type ECSs are defined by
$\ket{ECS^{\Phi \pm}_{\alpha}}_{BC} ={\cal N}^{\pm}_{\alpha} (\ket{\alpha}_{B} \ket{\alpha}_{C} \pm \ket{-\alpha}_{B} \ket{-\alpha}_{C})$ and
$\ket{ECS^{\Psi \pm}_{\alpha}}_{BC} ={\cal N}^{\pm}_{\alpha} (\ket{\alpha}_{B} \ket{-\alpha}_{C} \pm \ket{-\alpha}_{B} \ket{\alpha}_{C})$ \cite{Sanders12,Jeong02}.
\begin{figure}[t]
\includegraphics[width=8.0cm]{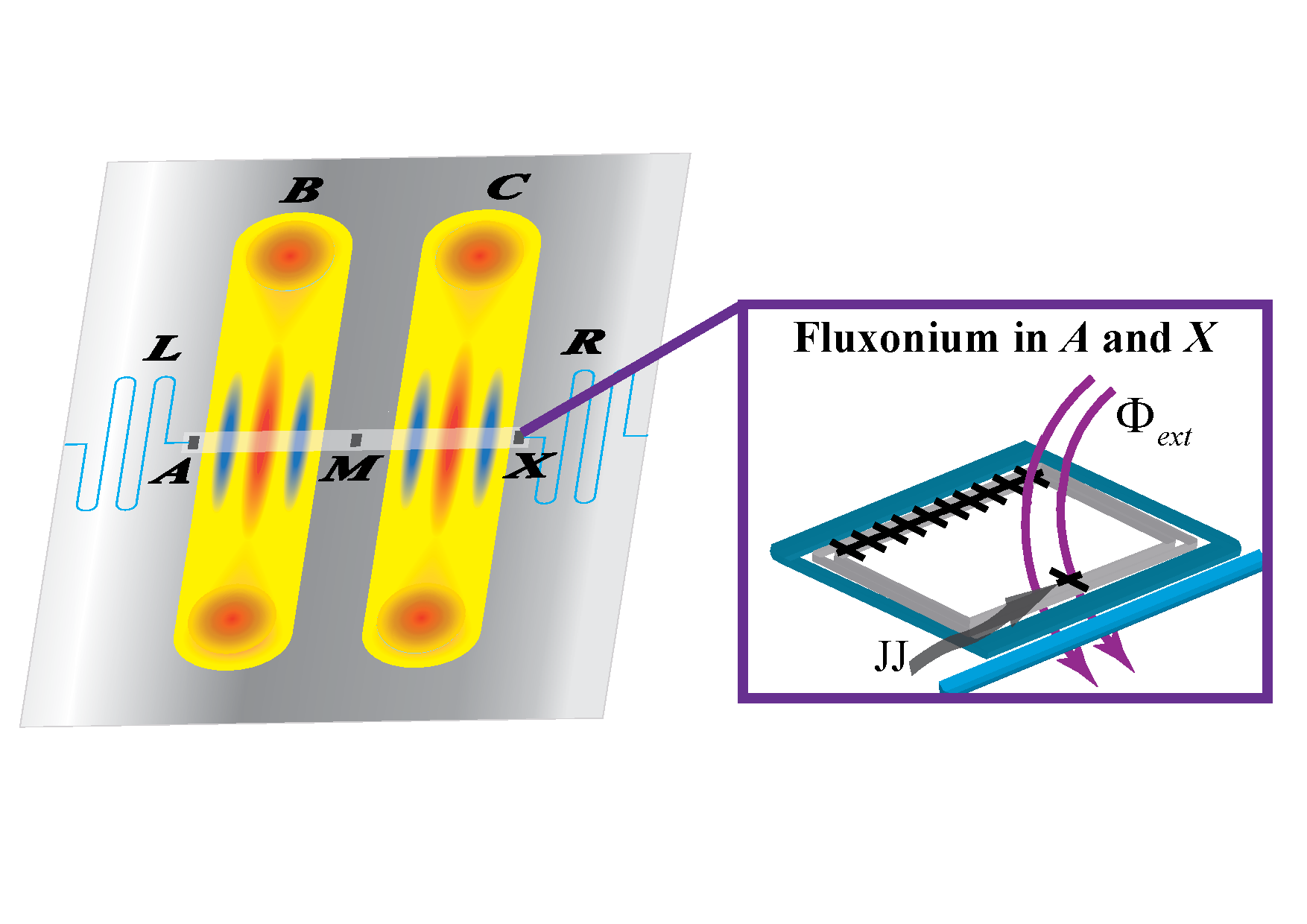}
\includegraphics[width=8cm]{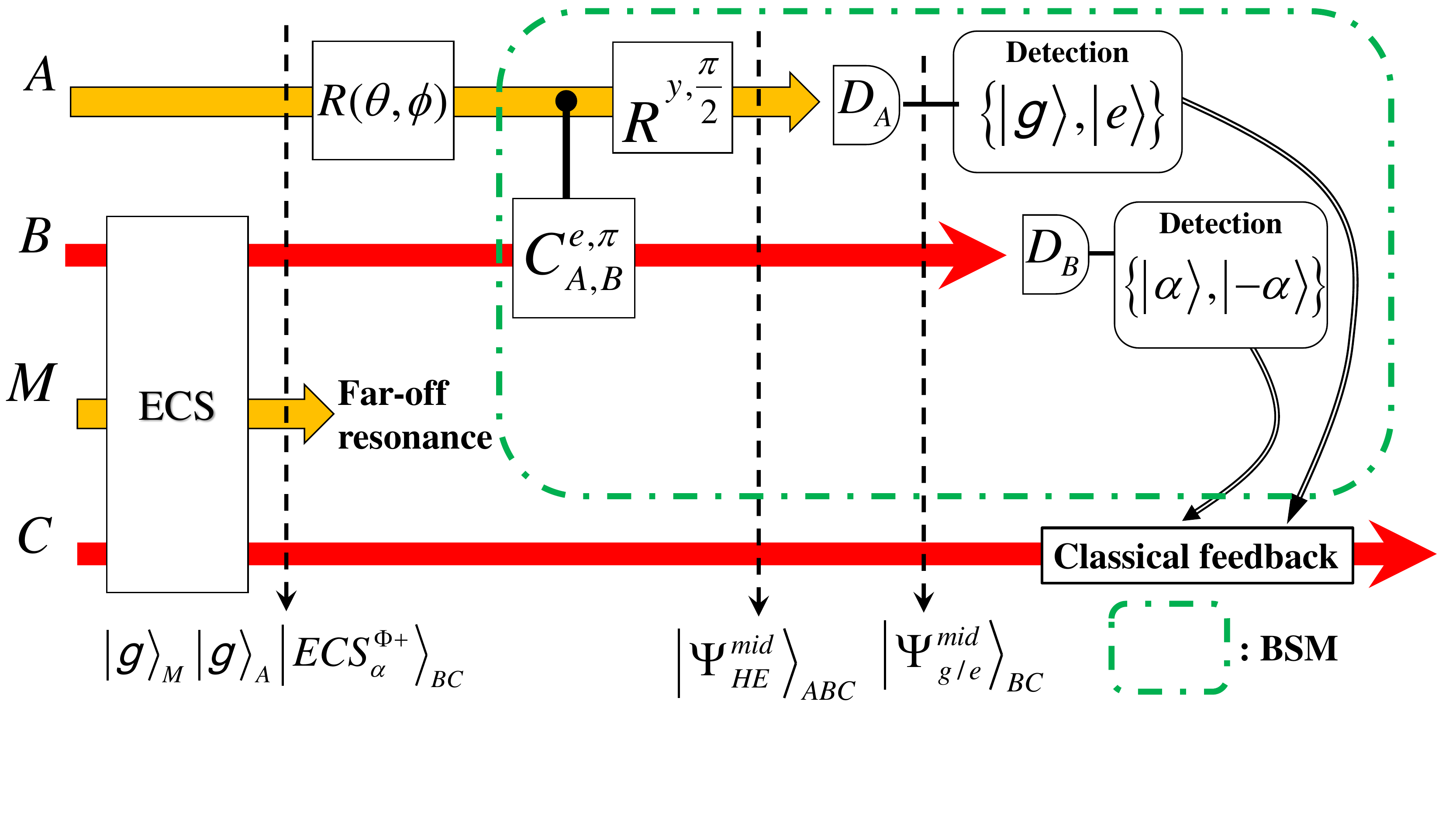}
\caption{(Top) An illustration of the architecture of a cavity-QED system named a two-cavity and three-qubit (2C3Q) architecture. The two high-Q cavities ($B$ and $C$) posses an ECS and the outer readout resonators ($L$ and $R$) are used for measurement of qubits and cavity fields. If the outer qubits ($A$ and $X$) are fluxonium, the self-Kerr distortion on the cavities might be reducible by an appropriate external flux $\Phi_{ext}$ (JJ: Josephson-Junction). (Bottom) Schematics of hybrid CV teleportation from an unknown superconducting state in $A$ to a CV cavity field in $C$. The states of $A$ and $M$ are superconducting qubits in orange lines while that of $B$ and $C$ are cavity fields in red lines. The transmon qubit $M$ is used for preparation of a ECS and decoupled enough cavity $C$ from $B$. The part of implementing the BSM scheme has been demonstrated very recently in \cite{Brian_arxiv}. }
\label{fig:qubit-teleportation}
\end{figure}

As similar to the DV-qubit teleportation, we begin with the unknown qubit given by $\ket{\psi}_A$ in a discretized two-level state, represented by ground and excited states of a superconducting qubit ($\ket{g} \equiv \ket{0}$ and $\ket{e} \equiv \ket{1}$), and the total initial state is given by
\begin{eqnarray}
&&\ket{\Psi^{tot}_{CV}}_{ABC} = \ket{\psi}_A \ket{ECS^{\Phi+}_{\alpha}}_{BC} , \label{Back_05} \\
&& ~~={\cal N}^{+}_{\alpha} \Big( \ket{\Phi^{+}_{HE}}_{AB} \otimes \openone  + \ket{\Phi^{-}_{HE}}_{AB} \otimes \tilde{\sigma}^{z}_C \nonumber \\
&&~~~~~~~~ + \ket{\Psi^{+}_{HE}}_{AB} \otimes \tilde{\sigma}^{x}_C + \ket{\Psi^{-}_{HE}}_{AB} \otimes \tilde{\sigma}^{x}_C \tilde{\sigma}^{z}_C \Big) \ket{\psi_{CV}}_C, \nonumber 
\end{eqnarray}
where 
the hybrid entangled states are 
\begin{eqnarray}
&& \ket{\Phi^{\pm}_{HE}}_{AB} = {1 \over \sqrt{2}}  (\ket{g}_{A} \ket{\alpha}_{B} \pm \ket{e}_{A} \ket{-\alpha}_{B}), \label{Back_07}  \\
&& \ket{\Psi^{\pm}_{HE}}_{AB} = {1 \over \sqrt{2}}  (\ket{g}_{A} \ket{-\alpha}_{B} \pm \ket{e}_{A} \ket{\alpha}_{B}), \label{Back_08} 
\end{eqnarray} 
and
\begin{eqnarray}
&& \ket{\psi_{CV}}_C = N^{a,b}_{\alpha} (a \ket{\alpha}_C + b \ket{-\alpha}_C),
\label{Back_06} 
\end{eqnarray} 
for $N^{a,b}_{\alpha}=1/\sqrt{1+2 \Re[b^*a] \exp(-2|\alpha|^2)}$ ($b^*$: a conjugate of $b$).

The hybrid BSM projects the state with the measurement set of $\{ \ket{\Phi^{\pm}_{HE}}_{AB} \bra{\Phi^{\pm}_{HE}}, \ket{\Psi^{\pm}_{HE}}_{AB} \bra{\Psi^{\pm}_{HE}} \}$. When Alice announces measurement outcomes, Bob obtains the teleported CV state as a generalized SCS and performs the classical feedback to recover the original unknown state in cavity mode $C$. Note that the final results clearly show the CV version of the original qubit upto pseudo-Pauli operators
\begin{eqnarray}
&& \tilde{\sigma}^{x}\,  \ket{\psi_{CV}}_C = N^{b,a}_{\alpha}  (b \ket{\alpha}_C + a \ket{-\alpha}_C), ~~~~ \label{Back_09} \\
&& \tilde{\sigma}^{z}\,  \ket{\psi_{CV}}_C = N^{a,-b}_{\alpha}   (a \ket{\alpha}_C - b \ket{-\alpha}_C).~~~~~ \label{Back_10} 
\end{eqnarray} 
In particular, even Schr\"odinger cat is given by $\ket{\psi_{CV}}_C = \ket{SCS^+_{\alpha}} = N^{1,1}_{\alpha} (\ket{\alpha}_C + \ket{-\alpha}_C)$ for $a=b$ while odd Schr\"odinger cat is given by $\ket{\psi_{CV}}_C = \ket{SCS^-_{\alpha}} = N^{1,-1}_{\alpha} (\ket{\alpha}_C - \ket{-\alpha}_C)$ for $a=-b$.

\subsection{Protocol implementation in a two-cavity and three-qubit (2C3Q) architecture }
\label{IIB}
We first describe how to implement this hybrid CV teleportation in circuit-QED. As shown in the top of Fig.~\ref{fig:qubit-teleportation}, we consider a specific architecture of two high-Q cavities, three qubits with two readout resonators at the edge (2C3Q), inspired by the existing experiment architecture \cite{Yale_big_cat}. The outer fluxonium qubits are in particular designed for reducing a self-Kerr nonlinearity in a cavity (see more details in Section \ref{self-Kerr}). In the circuit diagram of Fig.~\ref{fig:qubit-teleportation}, we assume that an qubit state and a channel state are initially prepared in the ground state of superconducting qubit $\ket{g}_A$ and in $\ket{ECS^{\Phi+}_{\alpha}}_{BC}$ in two cavity fields and the method of creating the ECS has already proposed in \cite{qcMAP13} using the middle transmon qubit $M$ (see a detailed circuit diagram of building the ECS in Fig.~\ref{fig:how2makeECS01}). According to the results in Ref. \cite{qcMAP13}, the fidelity of the generated ECS is estimated as 96\% in 190 ns under realistic defects given by self-Kerr and cross-Kerr effects. 

For preparation of an arbitrary qubit in $A$, a single-qubit operation $R(\theta,\phi)$ is applied on $\ket{g}_A$ given by
\begin{eqnarray}
\ket{\psi}_A && = R(\theta,\phi) \ket{g}_A = \cos {\theta \over 2} \ket{g}_A + e^{i \phi} \sin {\theta \over 2} \ket{e}_A.~~~ \label{Back_11}
\end{eqnarray} 
The transmon qubit frequency $\omega^T_M$ is far off from the cavity resonances to avoid/reduce a direct cross talk between two cavity fields. 
To perform the hybrid version of BSM on $A$ and $B$ (shown in a green box in Fig.~\ref{fig:qubit-teleportation}), two operations are firstly required such as a conditional phase gate between the superconducting qubit $A$ and the cavity state $B$ as well as a single-qubit rotation in $A$. A generalized conditional phase gate $C^{e,\varphi}$ is written by
 \begin{eqnarray}
C^{e,\varphi}_{AB}  && = e^{i\varphi \ket{e}_{A}\bra{e}\,\hat{n}_B} = \ket{g}\bra{g} \otimes  \openone +   \ket{e}\bra{e} \otimes e^{i\varphi \, \hat{n}_B},~~~~ \label{Back_12} 
\end{eqnarray} 
where $\hat{n} = \hat{a}^{\dag} \hat{a}$, and $R^{y,{\pi \over 2}}={1\over \sqrt{2}} (\openone+i \sigma^{y})$ .
After the single-qubit operation $R^{y,{\pi \over 2}}_A$ and $\varphi=\pi$, the total state is equal to
\begin{eqnarray}
 \ket{\Psi^{mid}}_{ABC} &&= \left[ \left( R^{y,{\pi \over 2}}_A \otimes \openone_B \right) \otimes C^{e,\pi}_{AB}  \right] \ket{\Psi^{tot}_{CV}}_{ABC}. ~~~~ \label{Back_13}
\end{eqnarray} 
Note that the operation of $R^{y,{\pi \over 2}}_A$ transfers $\ket{g} \rightarrow \ket{-}= (\ket{g}-\ket{e})/\sqrt{2}$ and  $\ket{e} \rightarrow \ket{+}=(\ket{g}+\ket{e})/\sqrt{2}$. The combination of these operations $\left( R^{y,{\pi \over 2}}_A \otimes \openone_B \right) \otimes C^{e,\pi}_{AB}$ makes the Bell states ($\ket{\Phi^{\pm}_{HE}}_{AB}$ and $\ket{\Psi^{\pm}_{HE}}_{AB}$) into four product states ($\ket{g}_A \ket{\pm \alpha}_B$ and $\ket{e}_A \ket{\pm \alpha}_B$). 

What Alice needs is now sequential detections on the state of $A$ and $B$ in the basis sets of $\{\ket{g}_A,\ket{e}_A \}$ and $\{\ket{\alpha}_B,\ket{-\alpha}_B \}$ through the low-Q resonator $L$. In Fig.~\ref{fig:qubit-teleportation}, two measurements are independently performed in the superconducting qubit ($\{\ket{g}_A,\ket{e}_A\}$) first and the cavity field ($\{\ket{\alpha}_B,\ket{-\alpha}_B \}$) later. After reading the qubit state in $A$ is $\ket{g}_A$ or $\ket{e}_A$, the CV-qubit measurement can be performed in $\{\ket{\alpha}_B,\ket{-\alpha}_B \}$ by recycling the superconducting state collapsed in $A$ and the similar measurement technique has been very recently demonstrated in Ref.~\cite{Brian_arxiv}. An extra displacement operation $D(\alpha)$ on $\ket{\pm \alpha}_B$ could bring the better distinguishability of the CV state ($\ket{2\alpha}_B$ and $\ket{0}_B$) because its minimum requirement is to identify a vacuum state $\ket{0}_B$ conclusively. Once the qubit- and cavity-state measurements are at the level of single-shot measurement with high fidelity, the success probability of the hybrid BSM will be 1/4 in each outcome for $\alpha \gg 1$ (as same as the conventional teleportation) due to the orthogonality of four measurement outcomes in the BSM while non-orthogonal basis measurement might occur with the probability of smaller (or bigger) than 1/4 for small $\alpha$.

After the measurements, the final outcome state in mode $C$ might become one of the four CV states ideally (e.g., Eqs.~(\ref{Back_06}) to (\ref{Back_10})) and Bob obtains one of the four states given by $ \ket{\psi^{fin}_{g/e,\pm\alpha}}_C \propto \big[ \bra{g/e}_A \bra{ \pm \alpha}_B\big] \, \ket{\Psi^{mid}}_{ABC}$ such that
\begin{eqnarray}
\ket{\psi^{fin}_{g,\alpha}}_C &&= N_{\alpha} \left( \cos {\theta \over 2} \ket{\alpha}_{C} + e^{i \phi} \sin {\theta \over 2} \ket{-\alpha}_{C} \right),   \label{Back_14}~~~ 
\end{eqnarray} 
and $\ket{\psi^{fin}_{g,-\alpha}} = \tilde{\sigma}^{x} \ket{\psi^{fin}_{g,\alpha}}$, 
$\ket{\psi^{fin}_{e,\alpha}} = \tilde{\sigma}^{z} \ket{\psi^{fin}_{g,\alpha}}$, and 
$\ket{\psi^{fin}_{e,-\alpha}} =  \tilde{\sigma}^{x} \tilde{\sigma}^{z} \ket{\psi^{fin}_{g,\alpha}}$. 
To verify the teleported state (pseudo single-qubit rotated) in mode $C$, the qubit $X$ and the most right low-Q resonator $R$ will be used for performing a Wigner function plot of the cavity state (see the top of Fig.~\ref{fig:qubit-teleportation}). Additionally, the unknown superconducting qubit state can be recovered in CV qubit in mode $C$ through the pseudo Pauli operator, which can be performed by a qcMAP gate with the superconducting qubit $X$ \cite{qcMAP13}.

\begin{figure}[t]
\includegraphics[width=8.5cm]{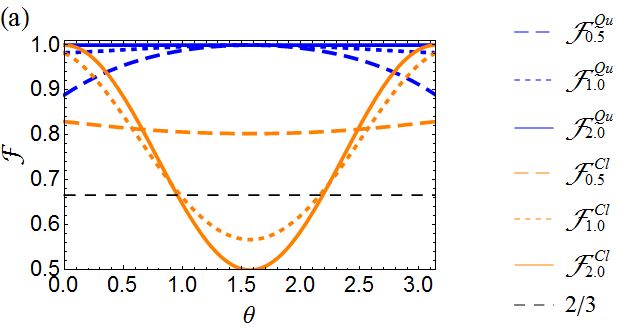}
\includegraphics[width=8.2cm]{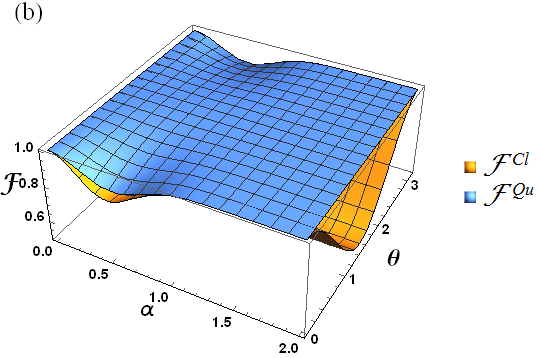}
\caption{Teleportation fidelities ${\cal F}^{Cl}_{\alpha}$ in Eq.~(\ref{classic_mix03}) and ${\cal F}^{Qu}_{\alpha}$ in Eq.~(\ref{F_QU01}) for $\phi=0$ and $\ket{g}_A$. (a) The orange curve shows that the classical teleportation fidelity cannot excess 1/2 for $\ket{\pm}_A$ ($\theta=\pi/2$) with $\ket{ECS^{\Phi+}_{2.0}}_{BC}$ while the fidelity of quantum teleportation is always 1. (b) ${\cal F}^{Qu}_{\alpha}$ (blue surface) is always better than ${\cal F}^{Cl}_{\alpha}$ with respect to $\alpha$ and $\theta$. For example, ${\cal F}^{Cl}_{2.0}$ is approximately equal to the fidelity of DV classical teleportation given by$\cos^4 {\theta \over 2} + \sin^4 {\theta \over 2}$.}
\label{fig:mixture_fidelity01}
\end{figure}

\subsection{Fidelity of hybrid teleportation}
We consider now the teleportation fidelity which will be determined by the quantumness of the channel state in a teleportation scheme when the BSM is ideal. If the channel suffers decoherence before the BSM, it should be described in mixed states. Based on the criteria of successful quantum teleportation in DV qubits, the average fidelity of a teleported state needs in theory to be higher than 2/3 to claim the validity of using a quantum channel because maximally correlated classical states (i.e., $\rho^{mix}_{BC} = (\ket{0}_{B}\bra{0}\otimes\ket{0}_{C}\bra{0} + \ket{1}_{B}\bra{1} \otimes \ket{1}_{C}\bra{1})/2$) can be used for performing classical teleportation upto the average fidelity 2/3 \cite{Popescu94}. Full CV teleportation, however, proposes the different criterion that the average fidelity larger than $1/2$ shows the nonclassicality of a teleportation channel \cite{Sam_Braunstein} since a classical channel (i.e., two coherent states ($\ket{\alpha}_B\ket{\alpha}_C$) produces the teleportation fidelity 1/2. This issue might be originally caused by the definition of quantumness and nonclassicality. 

In our quantum teleportation, the quantity of the fidelity relies on the degree of decoherence as well as the nonorthogonality given by the initial size of $\alpha$ of the channel state. For comparison with the fidelity of hybrid quantum teleportation, we define the fidelity of DV classical teleportation ${\cal F}^{Cl}$ with respect to the angle $\theta$ in the unknown state. For DV teleportation, ${\cal F}^{Cl} = \cos^4 (\theta / 2) + \sin^4 (\theta / 2)$ with the unknown state ($\phi = 0$) and the classically correlated channel  $\rho^{mix}_{BC}$. Thus, to claim that our CV channel is a nonclassical (or quantum) channel, the fidelity of hybrid teleportation should be described by than ${\cal F}^{Qu}$, which is better than ${\cal F}^{Cl}$ with parameter $\alpha$. For example, we compare the fidelity of hybrid teleportation with that of classical teleportation in the outcome of $\ket{g}_{A}\ket{\alpha}_B$ and explain which experimental condition would show a clear distinction between quantum and classical cases. 

Because the initial state is in DVs and the teleported state is in CV in our teleportation, we define the teleportation fidelity between a teleported state $\ket{\tilde{\psi}^{fin}_{g/e,\pm \alpha}}$ and an expected CV state  $\ket{\psi^{fin}_{g/e,\pm \alpha}}$  given by
\begin{eqnarray}
{\cal F}^{Qu}_{g/e,\pm \alpha} =\left| {}_C \Big< \psi^{fin}_{g/e,\pm \alpha} \Big| \tilde{\psi}^{fin}_{g/e,\pm \alpha} \Big>_C \right|^2 = \left| \tilde{N}_{\alpha} {N}_{\alpha} {\cal W}\right|^2 ,~~~~~
\label{F_QU01}
\end{eqnarray}
where ${\cal W} = 1 + 4 e^{-2|\alpha|^2} \cos \phi \cos {\theta \over 2} \sin{\theta \over 2} + e^{-4|\alpha|^2}$. For example, if Alice obtains the outcome state $\ket{g}_A \ket{\alpha}_B$, the teleported state is given by
$ \ket{\tilde{\psi}^{fin}_{g ,+ \alpha}} = \tilde{N}_{\alpha} ( \tilde{a} \ket{\alpha} + \tilde{b} \ket{-\alpha})$ for $\tilde{a} = \cos {\theta \over 2} + e^{-2|\alpha^2|} e^{i \phi} \sin {\theta \over 2}$ and
$\tilde{b} = e^{-2|\alpha^2|} \cos {\theta \over 2} + e^{i \phi} \sin {\theta \over 2}$. 
To claim that our hybrid teleportation has been performed through a nonclassical channel, we need to show that the hybrid teleportation fidelity exceeds the fidelity with a classically correlated state (as a classical channel) given by 
\begin{eqnarray}
 \rho^{mix,\alpha}_{BC} &=& {1 \over 2} (\ket{\alpha}_{B}\bra{\alpha}\otimes\ket{\alpha}_{C}\bra{\alpha} \nonumber \\
&& ~~+ \ket{-\alpha}_{B}\bra{-\alpha} \otimes \ket{-\alpha}_{C}\bra{-\alpha}), \label{classic_mix01}
\end{eqnarray}
which can be understood as the state suffering a decoherence from the ideal ECS channel.
Thus, the fidelity between $\ket{\psi^{fin}_{g,\alpha}}_C$ and a classically teleported state $\rho^{Cl}_C$ through $ \rho^{mix,\alpha}_{BC}$ is given by 
\begin{eqnarray}
{\cal F}^{Cl} ={}_{C} \bra{\psi^{fin}_{g, \alpha}} \rho^{Cl}_C \ket{\psi^{fin}_{g,\alpha}}_C.
\label{classic_mix03}
\end{eqnarray}
As explained above, ${\cal F}^{Cl}$ is approximately equal to $\cos^4 {\theta \over 2} + \sin^4 {\theta \over 2}$ for large $\alpha$, which is also obtained by DV classical teleportation with $\ket{\psi}_A$ and  $\rho^{mix}_{BC}$.
In contrast to DV- and CV-only teleportations, the fidelities decrease not only with the decoherence of the ECS channel but also with the size of $\alpha$ in the coherent-state representation. 
For example, for small $\alpha$, $\ket{ECS^{\Phi+}_{\alpha}}_{BC}$ tends to behave similar to two vacuum states but still maintains the superposition between $\ket{\alpha}\ket{\alpha}$ and $\ket{-\alpha}\ket{-\alpha}$. 

We here examine the fidelity characteristics with $\phi=0$ and the outcome of $\ket{g}_A\ket{\alpha}_B$. For example, the classically teleported state has lost the coherence of the unknown state and is given by
\begin{eqnarray}
\rho^{Cl}_{C} = {\cal M} \left( f_{+} \ket{\alpha}_C \bra{\alpha} + f_{-} \ket{-\alpha}_C \bra{-\alpha} \right),
\label{classic_mix02}
\end{eqnarray}
where $f_{+}= \cos^2 {\theta \over 2} + \sin^2 {\theta \over 2} e^{-4\alpha^2} + \cos {\theta \over 2} \sin {\theta \over 2} e^{-2\alpha^2}$ and $f_{-}= \cos^2 {\theta \over 2} e^{-4\alpha^2} + \sin^2 {\theta \over 2} + \cos {\theta \over 2} \sin {\theta \over 2} e^{-2\alpha^2}$. Note that $\rho^{Cl}_{C} \approx  \cos^2 {\theta \over 2}  \ket{\alpha}_C \bra{\alpha} +\sin^2 {\theta \over 2}  \ket{-\alpha}_C \bra{-\alpha}$ for $\alpha > 1$. 

As shown in the top of Fig.~\ref{fig:mixture_fidelity01}, the hybrid teleportation fidelity ${\cal F}^{Qu}_{\alpha} \ge {\cal F}^{Cl}_{\alpha}$ for fixed $\alpha$ overall. The fidelity of the hybrid teleportation needs to meet the criteria of CV teleportation fidelity given by ${\cal F}^{Cl}$ (orange lines in the top figure). For large $\alpha$, the curves show that ${\cal F}^{Cl}$ is far less than 1 while ${\cal F}^{Qu} \approx 1$. In particular, the value of ${\cal F}^{Cl}$ becomes 1/2 at around $\theta=\pi/2$ \cite{Sam_Braunstein}. The reason of ${\cal F}^{Cl} \approx 1$ for large $\alpha$ with $\theta=0$ is that the classical teleportation also works well if the unknown state is in $\ket{g}_A$ or $\ket{e}_A$ as a classical bit. In the bottom of Fig.~\ref{fig:mixture_fidelity01}, if $\alpha \approx 0$, $\ket{ECS^{\Phi+}_{\alpha}}$ and $\rho^{mix}_{BC}$ both become a vacuum and two fidelities reaches 1. 

Interestingly, for $\theta=0$ (or $\pi$) and $\alpha \approx 0.5$, the hybrid teleportation fidelity is far less than 1 because of the effect of the nonorthogonal measurement in $\{\ket{\alpha}_B,\ket{-\alpha}_B \}$. However, the fidelity of hybrid quantum teleportation is always the unity (${\cal F}^{Qu}_{g,\alpha}=1$) for $\theta=\pi/2$ and any size of $\alpha$ because the measurement outcomes of $\ket{g}_A \ket{\alpha}_B$ and $\ket{g}_A \ket{-\alpha}_B$ bring the identical outcome as $N^{1,1}_{\alpha} (\ket{\alpha}_C + \ket{-\alpha}_C)$. Thus, the issue of nonorthogonal measurement given by small $\alpha$ does not affect on the fidelity at $\theta = \pi/2$. Therefore, this hybrid quantum teleportation might be able to show a clear advantage from the equally superposed input state $\ket{\pm}_A$ to be teleported in even/odd Schr\"odinger cat states while $\ket{g}_A$ and $\ket{e}_A$ give the same amount of the fidelities for both classical and quantum teleportation. Therefore, if ${\cal F}^{Qu}_{g/e,\pm \alpha} > {\cal F}^{Cl}$, it is shown that the hybrid teleportation is performed through a nonclassical channel. 

\section{Additional applications in 2C3Q architecture}
\label{III}
We here propose two additional applications for engineering hybrid quantum information processing based on the same architecture shown in the top of Fig.~\ref{fig:qubit-teleportation}. First, a verification of ECSs can be performed by measuring cavity fields in a cavity-state measurement setup and this scheme will provide us an efficient method of quantum state tomography for specific entangled CV states. Second, a single-qubit error in entangled CV qubits can be monitored by entangling and measuring outer ancillary qubits. This is equivalent to a non-destructive syndrome measurement, which is useful for robust quantum information processing in circuit-QED. 
\begin{figure}[t]
\includegraphics[width=8cm,trim=0cm -1.5cm 0cm 0cm]{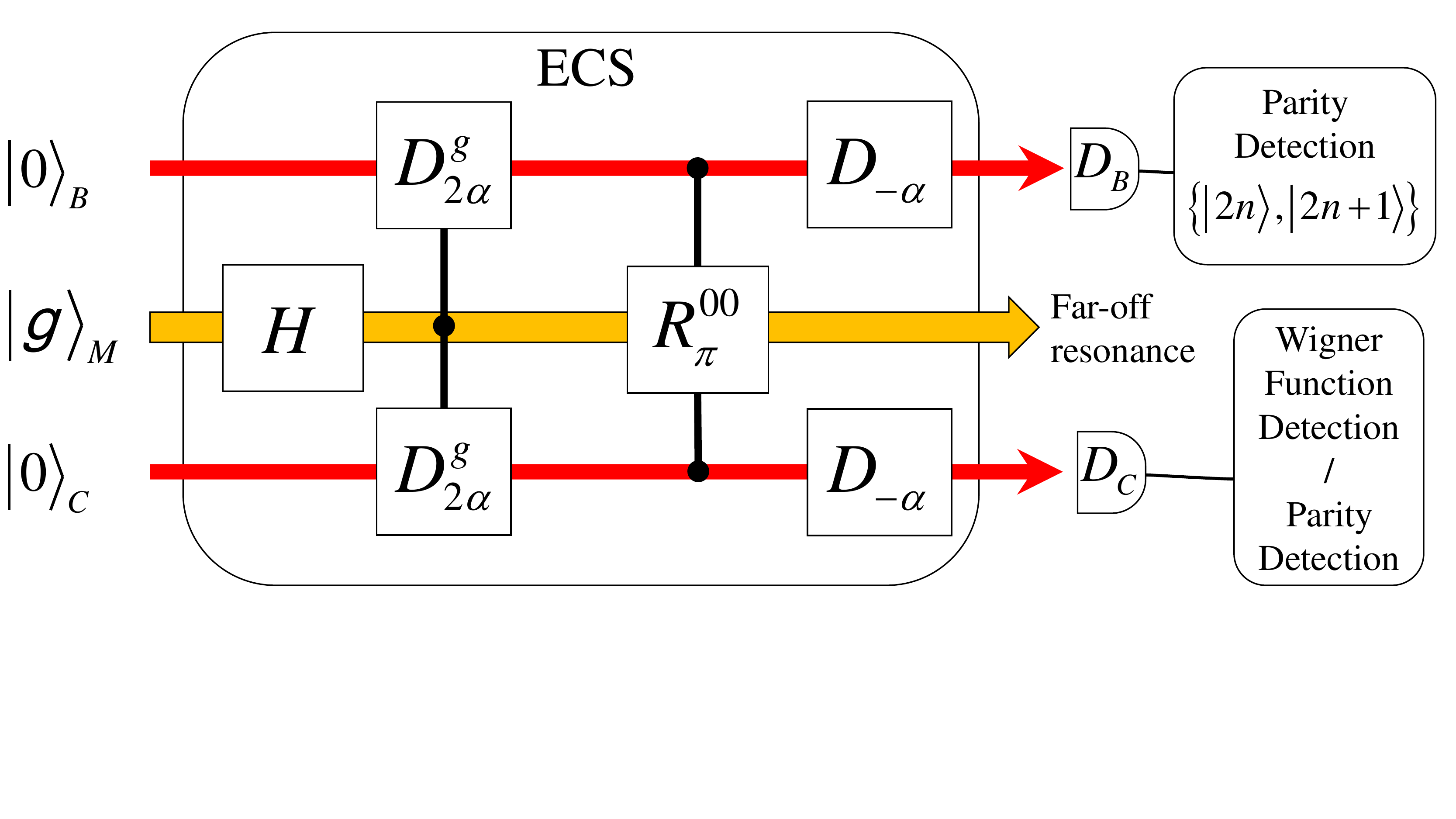}
\vspace{-1.cm}
\caption{Circuit diagram for creating and verifying an ECS \cite{qcMAP13}. $D^{g}_{2\alpha}$ is a conditional displacement with $2\alpha$ while $D_{-\alpha}$ does unconditional one with $-\alpha$. $R^{00}_{\pi}$ makes a $\pi$-flip operation of the qubit $M$ at a vacuum state in modes $B$ and $C$. $D_B$ and $D_C$ indicate detectors for cavity states.}
\label{fig:how2makeECS01}
\end{figure}

\subsection{Verification scheme for ECSs} 
Because CV states in principle have infinite dimension, this brings a difficulty to perform a conventional quantum tomography for CV states generally (e.g., measurement of CV states in all Fock states). However, an ECS can be verified with two sets of measurement schemes if the orthogonality between $\ket{\alpha}$ and $\ket{-\alpha}$ are large enough (e.g. $\alpha>1.5$). One is a measurement setup in the basis-state set  $\{\ket{\alpha}, \ket{-\alpha} \}$ and the other is a parity measurement (or a measurement of Wigner functions) of CV states. For ECS verification, we here use the same architecture for hybrid teleportation as shown in Fig.~\ref{fig:qubit-teleportation}. If $\ket{ECS^{\Phi +}_{\alpha}}_{BC}$ is prepared in the cavities of $B$ and $C$, one measures the cavity states individually connected to outer superconducting qubits ($A$ and $X$) and two outer resonators ($L$ and $R$). 

The basic idea of our verification scheme is inspired from a stabilizer formalism on a Bell state. To verify a Bell state  $\ket{\Phi^+}=(\ket{00}_{BC}+\ket{11}_{BC})/\sqrt{2}$, one only needs to obtain the expectation values of $\sigma^z_B \sigma^z_C$ and $\sigma^x_B \sigma^x_C$ on $\ket{\Phi^+}$ because $\ket{\Phi^+}$ is the only state which always provides +1 eigenvalue for the two sets of Pauli operators given by
\begin{eqnarray}
\label{syndrom01}
\bra{\Phi^+}(\sigma^z_B \sigma^z_C)\ket{\Phi^+}=\bra{\Phi^+}(\sigma^x_B \sigma^x_C)\ket{\Phi^+}=1.
\end{eqnarray}

In order to perform a verification scheme on the even ECS, we adopt the known scheme of creating an ECS presented in \cite{qcMAP13} (see the box of the circuit diagram in Fig.~\ref{fig:how2makeECS01}). After the initial state of $\ket{g}_M$ with two vacuum states in modes $B$ and $C$, a conditional displacement operation $D^{g}_{2\alpha}$ through qubit M creates entanglement among the qubit and two cavity fields. As shown in the details in Ref.~\cite{qcMAP13}, a conditional qubit rotation $R^{00}_{\pi}$ disentangles the cavity-state channel from the qubit and the entangled CV state becomes the form of a maximally entangled state proportional to $\ket{0}_B \ket{0}_C +\ket{2\alpha}_B \ket{2\alpha}_C$. After the unconditional displacement operation $D_{-\alpha}$, The outcome state is finally given by $\ket{ECS^{\Phi +}_{\alpha}}_{BC}$. To measure the amount of entanglement, a Bell-type nonlocality test can be used by observing Wigner functions of two cavity states \cite{qcMAP13}, however, an efficient scheme of verification on the prepared state is a different approach from that how much the amount of entanglement the state has. 

In this ECS verification, two independent detections are first performed in the basis-state set $\{\ket{\alpha}, \ket{-\alpha} \}$ in both modes $B$ and $C$. This measurement results can provide the expectation value of $\tilde{\sigma}^z_C \tilde{\sigma}^z_C$ on $\ket{ECS^{\Phi +}_{\alpha}}_{BC}$ and show the correlated measurement outcomes of two cavity states such as $\ket{\alpha}_B \ket{\alpha}_C$ or $\ket{-\alpha}_B \ket{-\alpha}_C$ if $\ket{\alpha}$ is enough orthogonal to $\ket{-\alpha}$. As we mentioned in the scheme of the hybrid BSM in Section \ref{IIB}, an additional displacement operation might give an easier measurement scheme given by detecting a vacuum state $\ket{0}$ conclusively. Thus, the outcomes will be identical in both modes $B$ and $C$ if the prepared state is given by $\ket{ECS^{\Phi +}_{\alpha}}_{BC}$.  

Even if the outcomes are perfectly correlated in the $\tilde{\sigma}^z$ measurement, it does not however provide sufficient information for the verification of the CV state because the perfect correlation might come from the classically correlated state $\rho^{mix,\alpha}_{BC}$ but not the quantum correlated state $\ket{ECS^{\Phi +}_{\alpha}}_{BC}$. To distinguish them, a parity measurement is required on one (or both) cavity field(s) as the pseudo Pauli-$x$ operation. This parity measurement scheme has been tested in the similar superconducting circuits \cite{Yale_parity}. For example, the parity measurement in mode $C$ is equivalent to project the state of $C$ onto the basis-state set of even/odd SCSs and forces to collapse the cavity state of $B$ into an even/odd SCS such that
\begin{eqnarray}
\ket{ECS^{\Phi +}_{\alpha}}_{BC}  && \propto \ket{SCS^+_{\alpha}}_B \ket{SCS^+_{\alpha}}_C + \ket{SCS^-_{\alpha}}_B \ket{SCS^-_{\alpha}}_C,\nonumber \\
\label{Parity_check01}
\end{eqnarray}
where even SCSs have the sum of even photon-number states $\ket{SCS^+_{\alpha}} =\sum_{n} d_{2n} \ket{2n}$ and odd SCSs do that of odd photon-number states $\ket{SCS^-_{\alpha}}=\sum_{n} d_{2n+1} \ket{2n+1}$. In other words, the outcome state in mode $B$ has to have a fringe pattern in Wigner function distribution in $B$ after the parity measurement in $C$ \cite{Yale_parity} because the outcome of even (odd) parity brings an even (odd) SCS in mode $B$. Thus, the perfect correlation of the parity measurement outcomes occurs only if the prepared state is the ECS. On the other hand, this parity measurement on one of the classically correlated state in Eq.~(\ref{classic_mix01}) will provide a fully mixed state and no fringe patten in mode $B$ given by
\begin{eqnarray}
\rho^{mix}_{BC} ={1 \over 2} (\ket{\alpha}_{B}\bra{\alpha} + \ket{-\alpha}_{B}\bra{-\alpha}) \otimes \sum_{n} w_{n} \ket{n}_C\bra{n}.~~~~~~
\end{eqnarray}
Therefore, two measurement sets of pseudo Pauli operators can verify the state of  $\ket{ECS^{\Phi +}_{\alpha}}_{BC}$ in two cavities.
\begin{figure}[t]
\includegraphics[width=8cm,trim=0cm 0cm 0cm 0cm]{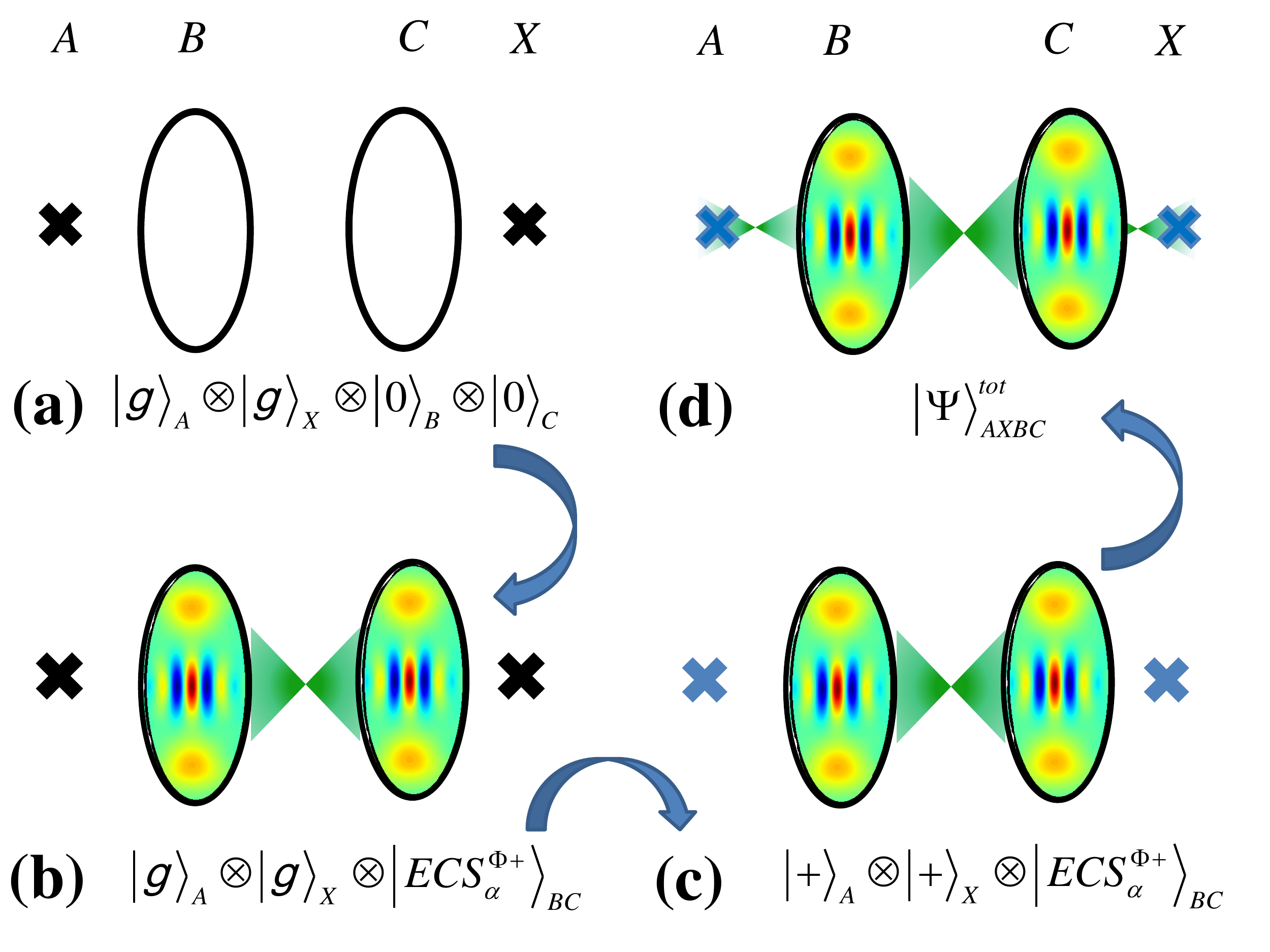}
\caption{(a) The initial state is prepared in $\ket{g}_A \ket{g}_X \ket{0}_B \ket{0}_C$. (b) The ECS is created by the scheme in Fig.~\ref{fig:how2makeECS01}. After two Hadamard operations in superconducting qubits of $A$ and $X$ in (c), two entangling gates between the superconducting qubit $A$ ($X$) and the CV qubit $B$ ($C$) given by$C^{e,\pi}_{AB} \, C^{e,\pi}_{CX}$ to create the four-partite entangled state in (d).}
\label{fig:PauliX}
\end{figure}

\subsection{Single-qubit Pauli-$x$ gate on ECSs}
A set of one- and two-qubit gates and Pauli-gates are simplest and essential gates for universal quantum computing. As shown in the hybrid teleportation scheme, the final state before classical feed-forward is indeed a single CV-qubit operated state (see below in Eq.~(\ref{Back_14})). We here present a different and deterministic scheme of a pseudo Pauli-$x$ gate on an ECS. In DV qubits, a Pauli-$x$ gate is known as the quantum NOT gate performing an operation between $a\ket{0}+b\ket{1} \leftrightarrow b\ket{0}+ a\ket{1}$ and it is equivalent to operate the gate between $a\ket{\alpha}+b\ket{-\alpha} \leftrightarrow b\ket{\alpha}+ a\ket{-\alpha}$. Since the verification of ECSs guarantees the quality of a ECS preparation, we are ready for CV quantum information processing in superconducting circuits. In the 2C3Q architecture, a repeat-until-success $\tilde{\sigma}^x$ gate can be performed by repeating entangling and measuring outer superconducting qubits through resonators within decoherence time. It is similar to the above verification scheme because both also rely on the stabilizer formalism given in Eq.~(\ref{syndrom01}). The key difference is however the fact that ancillary qubits, additionally entangled with the ECS, provide information of the CV qubits without destroying the prepared ECSs. Thus, this scheme can be in general applicable for the $\tilde{\sigma}^x$ operation in multipartite CV entangled states. For the syndrome detection without a CV state collapse, we create a four-partite hybrid entangled state between the outer superconducting qubits ($A$, $X$) and the ECS in $B$ and $C$, and then, the parity of the outer superconducting qubits is detected in the measurement set of $\{ \ket{g}, \ket{e}\}$. 

As the schematic protocol is depicted for building a specific four-qubit entangled state in Fig.~\ref{fig:PauliX}, (a) we begin with four separable states in two superconducting qubits and two cavities such as $\ket{g}_A \ket{g}_X \ket{0}_B \ket{0}_C$. (b) The method of creating the ECS is performed by Fig.~\ref{fig:how2makeECS01} and the state is prepared in $\ket{g}_A \ket{g}_X \ket{ECS^{\Phi+}}_{BC}$. Note that we omit the superconducting qubit $M$ between two cavities here because the qubit is far-detuned from cavity frequency and does not participate in the operations after creating the ECS. (c) A Hadamard operation are performed in the superconducting qubits resulting in $\ket{+}_A \ket{+}_X \ket{ECS^{\Phi+}}_{BC}$ (e.g., $\ket{+}=(\ket{g}+\ket{e})/\sqrt{2}$). (d) After the entangling operation of $C^{e,\pi}_{AB} \, C^{e,\pi}_{CX}$, the four-partite entangled state is equal to
\begin{eqnarray}
\ket{\Psi}^{tot}_{AXBC} &=& {1 \over \sqrt{2}} 
\Big( \ket{\Phi^+}_{AX} \ket{ECS^{\Phi+}_{\alpha}}_{BC} \nonumber \\
&& ~~~~~+ \ket{\Psi^+}_{AX} \ket{ECS^{\Psi+}_{\alpha}}_{BC} \Big).
\label{4QES01}
\end{eqnarray}
This state is known as a four-qubit Greenberger-Horne-Zeilinger state in two superconducting qubits and two cavity states because the state in Eq.~(\ref{4QES01}) is rewritten by
\begin{eqnarray}
\ket{\Psi}^{tot}_{AXBC} &=& {1 \over \sqrt{2}} 
\Big( \ket{++}_{AX} \ket{SCS^+_{\alpha}}_{B} \ket{SCS^+_{\alpha}}_{C}  \nonumber \\
&& ~~+\ket{--}_{AX} \ket{SCS^-_{\alpha}}_{B} \ket{SCS^-_{\alpha}}_{C}  \Big).
\label{4QES02}
\end{eqnarray}
Thus, the measurement outcomes of qubits $A$ and $X$ in $\{ \ket{g}, \ket{e}\}$ determine the two-cavity state in $\ket{ECS^{\Phi+}_{\alpha}}_{BC}$ or $\ket{ECS^{\Psi+}_{\alpha}}_{BC}$ in Eq.~(\ref{4QES01}) while the measurements in $\{ \ket{+}, \ket{-}\}$ brings a product state of two SCSs in modes $B$ and $C$ in Eq.~(\ref{4QES02}). 

For the repeat-until-success protocol, if the measurement outcomes are $\ket{g}$ (or $\ket{e}$) in both superconducting qubits, the two-cavity state is still kept in $\ket{ECS^{\Psi+}_{\alpha}}_{BC}$ while the cavity state is successfully collapsed into the desired state of $\ket{ECS^{\Psi+}_{\alpha}}_{BC}$ with different measurement outcomes in modes $A$ and $X$. Thus, the former outcome becomes the state depicted in Fig.~\ref{fig:PauliX}(b) and two entangling gates are performed again between $A$ and $B$ as well as $C$ and $X$ within coherence time in Fig.~\ref{fig:PauliX}(c, d). Finally, one of the cavity states is conditionally flipped from $\ket{\alpha}$ to $\ket{-\alpha}$ on $\ket{ECS^{\Phi+}_{\alpha}}_{BC}$.
\begin{figure}[t]
\includegraphics[width=7.5cm]{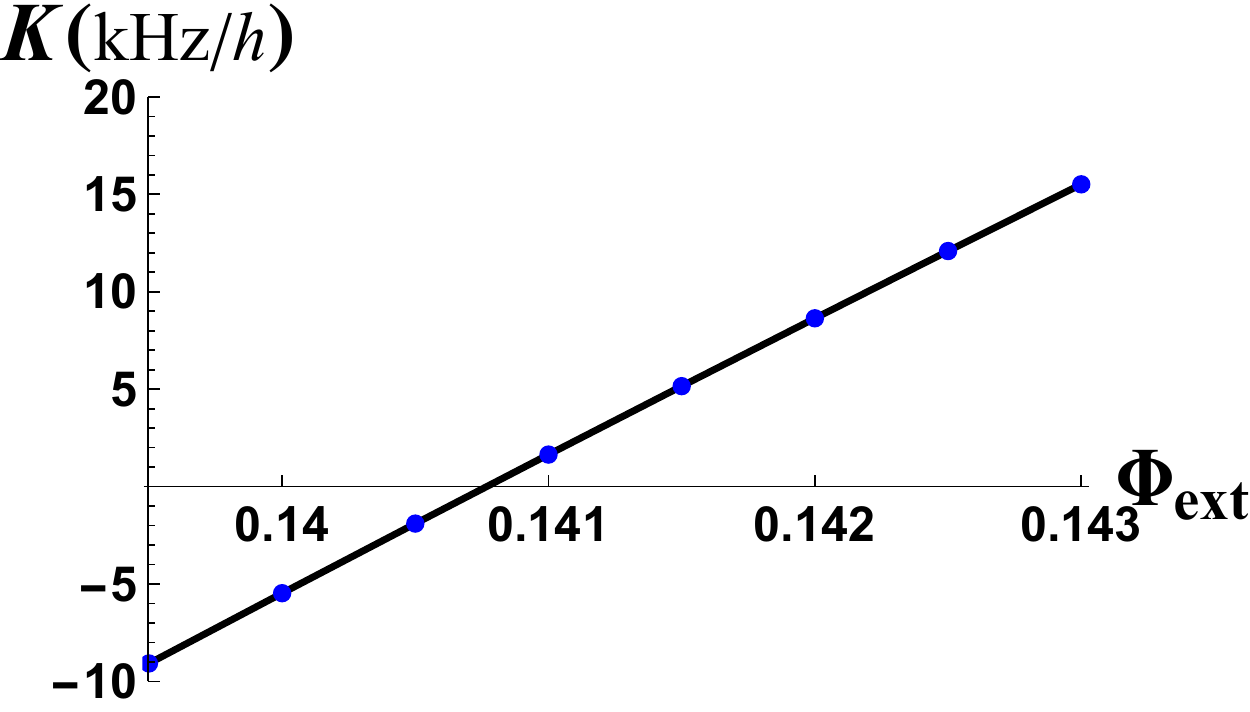}
\caption{It shows that the self-Kerr effect $K$ of a cavity almost linearly changes with respect to the external flux $\Phi_{ext}$ in a fluxonium.}
\label{fig:self-Kerr01}
\end{figure}

\section{Discussion and conclusion}
\subsection{Reduction of self-Kerr effects in a cavity field} 
\label{self-Kerr}
As pointed out the imperfect preparation of the ECS in Ref.~\cite{qcMAP13}, one of the major hurdles over the proposed schemes will be how to reduce a self-Kerr effect in a cavity induced by a nonlinearity of neighbouring superconducting qubits and a scheme of reducing self-Kerr interaction on a cavity will significantly diminish the distortion of the state during creating an ECS. This issue of self-Kerr detects will be continuously accumulated during longer operation time of quantum information processing. In the 2C3Q architecture, the key idea of reducing this distortion is that to utilize a different type of superconducting qubits, here which a fluxonium provides the capability of reversing the originally induced self-Kerr effect by the transmon qubit because it has a positive anharmonicity and hence an opposite influence on the cavity. The method of selective number-dependent arbitary phase gates (called SNAP gate) has been recently demonstrated in coherent states and could make additional reduction of the self-Kerr defects dynamically in the form of SCSs \cite{SNAP}. In contrast our method yields a reduction of self-Kerr by design and requires no further state mnipulation.

To demonstrate the principle of the reducing method in superconducting circuits, we examine a half of the 2C3Q system due to its symmetry of the full architecture and a cavity is sandwiched between a fluxonium and a transmon given by the Hamiltonian 
\begin{eqnarray}
\label{total_Ham01}
&& \hat{H}^{FCT} = \hat{H}_F + \hat{H}_T  + \hat{H}_C + \hat{H}_{FC} + \hat{H}_{CT} \\ 
&& =  \sum_{j,S} \omega^S_{j} \ket{j}_{S}\bra{j} + \omega^C \hat{a}^{\dag} \hat{a} +  \sum_{jk,S} \lambda^{S}_{jk} (\ket{k}_{S}\bra{j} \hat{a} + \ket{j}_{S}\bra{k} \hat{a}^{\dag}), \nonumber
\end{eqnarray}
for $S= F, T$ and $j<k$ ($\hbar =1$).
In order to estimate a single-photon Kerr effect \cite{1-photon_Kerr, EranPRB}, we examine the Kerr frequency $K$ in the effective Hamiltonian for cavity photons given by 
\begin{eqnarray}
\label{Effective_KerrHam}
\hat{H}^{\rm eff} = \tilde{\omega}^C  \hat{a}^{\dag} \hat{a} + {K \over 2} (\hat{a}^{\dag} \hat{a})^2.
\end{eqnarray}
The pseudo-photonic eigenstates and eigenvalues given by transmon and fluxonium ground states with 0, 1, and 2 photons in the cavity mode can be calculated given by $j,k=0,1,2$. 

Fig.~\ref{fig:self-Kerr01} shows that the effective self-Kerr effect in a cavity alters with respect to the external flux through a fluxonium.  For realistic parameters, we set up the cavity frequency as $\omega^C= 9.2$ GHz and the transmon energy levels $\omega^T_{j}$ and coupling strengths $\lambda^{T}_{jk}$ are given by $E^T_J=38$ GHz and $E^T_C=0.25$ GHz (see details in Ref.~\cite{Transmon_para}) while the fluxonium parameters are $E^F_L=0.5$ GHz, $E^F_J=8.5$ GHz and $E^F_C=3.0$ GHz \cite{Fluxonium_Ham}. For example, we obtain $K\approx -66.7$ kHz in a system of a transmon and a cavity with the absence of a fluxonium ($\omega^F_{j} = \lambda^{F}_{jk}=0$) while $K\approx 170.8$ kHz is given by the system of a fluxonium and cavity with $\omega^T_{j} = \lambda^{T}_{jk} =0$. If we include both superconducting qubits with the cavity commonly connected, the self-Kerr effect $K$ can be reduced. For example, if $|\Phi_{ext}|=0.141$, $\lambda^{F}_{01}\approx 0.038 $ GHz, $\lambda^{F}_{12}\approx 0.054 $ GHz, and $\lambda^{F}_{02}\approx 0.122 $ GHz \cite{Fluxonium_Ham} while $\lambda^{T}_{01}=0.10$ GHz, $\lambda^{T}_{12}\approx 0.141$ GHz, and $\lambda^{T}_{02}=0$. Then, the self-Kerr effect reduces $K\approx 1.64$ kHz (from $-66.7$ kHz in the system of a transmon and a cavity). Therefore, the architecture of the fluxonium-cavity-transmon can tune the self-Kerr effect in a cavity by design in circuit-QED systems.

\subsection{Conclusion}
We propose a teleportation scheme from a superconducting DV qubit to a microwave CV qubit in superconducting circuits. The proposed architecture of two cavities and three superconducting qubits is currently feasible with realistic parameters in the state-of-the-art platform of circuit-QED. The unknown state in a superconducting qubit is teleported via the ECS created between two cavities. The hybrid Bell measurement encodes the quantum information in the unknown qubit into a continuous-variable qubit in a cavity state. The teleportation fidelity in the hybrid scheme can confirm that the ECS channel is a nonclassical resource with respect to the size of $\alpha$. The same architecture is also beneficial for other CV quantum information processing for the schemes of verification and error-correction in the ECS channel. Finally, we presented a method of reducing a self-Kerr distortion in a cavity induced by two different superconducting qubits. 

Toward hybrid measurement-based quantum computing in circuit-QED, the capability of building a two CV-qubit gate between two cavities might be of essence in addition to single-qubit gates in superconducting circuits \cite{CVsingle-qubit}. For example, linear four-qubit hybrid cluster states will give a strength of one- and two-qubit gates which has been investigated in photonic measurement-based quantum computation \cite{Terry05}. To overcome errors in both superconducting and cavity qubits, we may need to build logical hybrid qubits with logical cluster states or to entangle higher-dimensional CV-qudits with superconducting qubits \cite{Andersen+Akira15, Cat_QEC}. For creating multi-partite ECSs, cross-Kerr interaction could be used in the multiple-cavity architecture joined by mediating qubits in order to keep the capability of CV-qubit operations in a dispersive regime. Furthermore, a full simulation of creating an ECS and of performing the hybrid quantum information processing in the 2C3Q architecture will be presented in later work \cite{Full_sim01}.

\section{Acknowlegements}
We would like to thank T. Spiller and B. Vlastakis for useful comments and T. Kim for assistance with graphics. EG acknowledges support from EPSRC (EP/L026082/1).

\end{document}